\documentclass[runningheads]{svmult}

\usepackage{makeidx}   
\usepackage{graphicx}  
\usepackage{subeqnar}  
\usepackage{multicol}  
\usepackage{physprbb}  
\makeindex             

\newcommand{\kms}{km s$^{\rm -1}$}
\newcommand{\kmskpc}{km s$^{\rm -1}$ kpc$^{\rm -1}$}
\begin{document}
\title*{VLT spectroscopy of globular clusters in the 
Sombrero galaxy}
%
%
%
%
\titlerunning{VLT spectroscopy of globular clusters in M\,104}
%
%
\author{Enrico V.Held\inst{1}
\and Alessia Moretti\inst{1,2}
\and Luciana Federici\inst{3}
\and Carla Cacciari\inst{3}
\and Luca Rizzi\inst{1}
\and Vincenzo Testa\inst{4}
}

\authorrunning{Enrico V. Held et al.}
%
%
\institute{
INAF, Osservatorio Astronomico di Padova, Padova, Italy
\and Dipartimento di Astronomia,
     Universit\`a di Padova,
     Padova, Italy
\and 
INAF, Osservatorio Astronomico di Bologna, Bologna, Italy
\and 
INAF, Osservatorio Astronomico di Roma, Monteporzio Catone, Italy
}
\maketitle              

\begin{abstract}
We have obtained intermediate-resolution VLT spectroscopy of 75
globular cluster candidates around the Sa galaxy M\,104 (NGC\,4594).
Fifty-seven candidates out to $\sim 40$ kpc in the halo of the galaxy
were confirmed to be bona-fide globular clusters, 27 of which are new.
A first analysis of the velocities provides only marginal evidence for
rotation of the cluster system.  From H$\beta$ line strengths, almost
all of the clusters in our sample have ages that are consistent,
within the errors, with Milky Way globular clusters. Only a few
clusters may be 1--2 Gyr old, and bulge and halo clusters appear
coeval.  The absorption line indices follow the correlations
established for the Milky Way clusters.  Metallicities are derived
based upon new empirical calibrations with Galactic globular clusters
taking into account the non-linear behavior of some indices (e.g.,
Mg2).  Our sample of globular clusters in NGC\,4594 spans a
metallicity range of $-2.13 < \mbox{[Fe/H]} < +0.26$ dex, and the
median metallicity of the system is [Fe/H]$ = -0.85$.  Thus, our data
provide evidence that some of the clusters have super-solar
metallicity. Overall, the abundance distribution of the cluster system
is consistent with a bimodal distribution with peaks at [Fe/H]$ \sim
-1.7$ and $-0.7$.  However, the radial change in the metallicity
distribution of clusters may not be straightforwardly explained by
a varying mixture of two sub-populations of red and blue clusters.
\end{abstract}

\section{Observations and reduction}
Multiobject spectroscopy of globular cluster candidates in the halo of
M\,104 was obtained using FORS1 at the ESO VLT in service mode in
April 2000. Five MOS setups were observed with exposure time $2 \times
50$ min each with the 600B grism and 1$^{\prime\prime}$ wide slitlets, 
yielding a resolution of $\sim 6$ \AA\ (FMHM).  All the available 19
movable slits were used, giving the highest priority to globular
clusters spectroscopically confirmed by Bridges and
coll. \cite{brid+97}.  In total 75 objects were observed, including
candidates visually selected on the basis of their shape and
magnitude. 
Since most targets were observed in more than one mask, the typical
target exposure time is 200 min. Long-slit spectra of Galactic
globular clusters spanning a wide range in metallicity were also
obtained for radial velocity and line strength calibration. 

The MOS spectra were extracted and wavelength calibrated using IRAF
standard tasks. The cross-correlation task FXCOR was employed to
measure radial velocities, using integrated spectra of Galactic
globular clusters as templates, and checking the velocity systematics
against synthetic spectra of simple stellar populations 
\cite{vazd99}.  In total 57 objects were confirmed to be
globular clusters in the Sombrero galaxy on the basis of their radial
velocities, 27 of which are new. 

\section{Metallicities and ages}

\begin{figure}[t]
\begin{center}
\includegraphics[width=.9\textwidth]{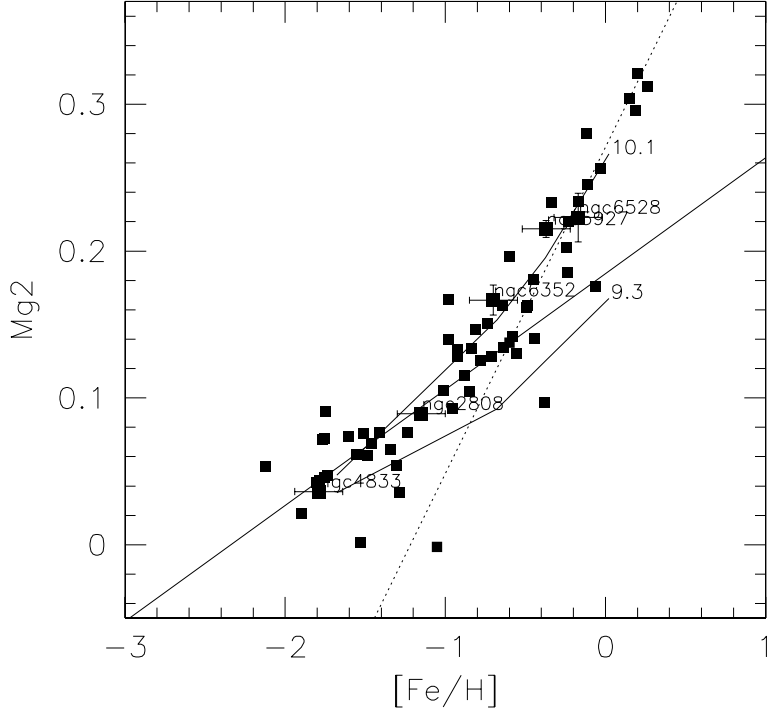}
\end{center}
\caption[]{The calibration of the Mg2 index against metallicity shows
a distinct non-linearity.  The labeled symbols with error bars
represent the Galactic globular cluster templates. Superimposed are
model curves from Bressan et al. \cite{bres+96} for two different
ages. The {\it solid line}
represents the empirical calibration of Brodie \& Huchra
\cite{brod+huch90}, valid up to [Fe/H]$ \sim -0.7$, while a change in
slope is seen for metal-rich clusters ({\it
dotted line}).  }
\label{f_heldF1}
\end{figure}

\begin{figure}[t]
\begin{center}
\includegraphics[width=.9\textwidth]{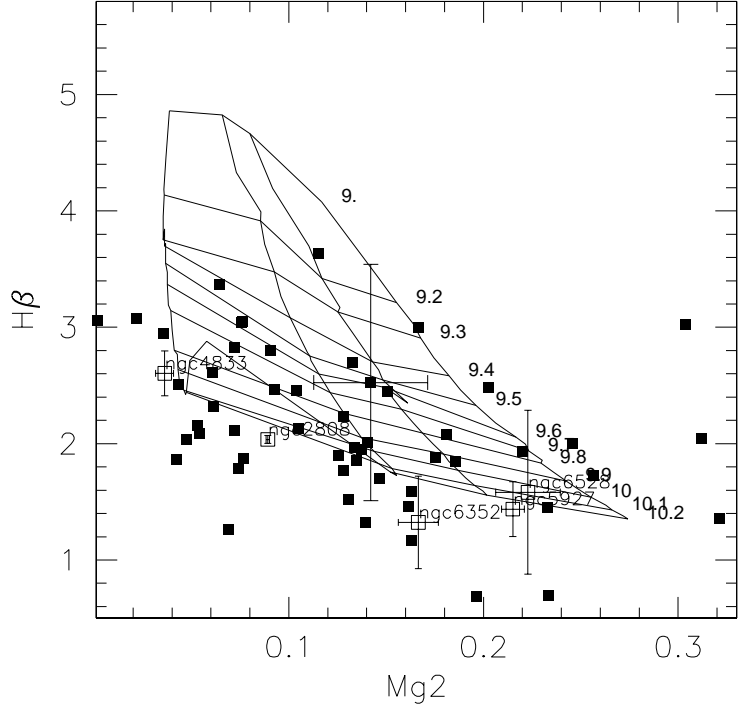}
\end{center}
\caption[]{The diagnostic diagram using H$\beta$ against Mg2 for
globular clusters around M\,104 ({\it filled squares}). Also shown are
the calibration data for Milky Way globular clusters ({\it open
squares} with error bars); overplotted is a model grid from Bressan et
al. \cite{bres+96}.  }
\label{f_heldF2}
\end{figure}

Absorption line indices were measured for all the confirmed clusters,
after convolving the spectra with an appropriate wavelength-dependent
kernel to match the resolution of the Lick/IDS system. The line
strengths were measured using the passbands defined by Worthey
et al. \cite{wort94}. The measurements of the template Galactic
globular clusters were compared with previous index data 
\cite{brod+huch90} to put our measurements on a standard
system.  Small offsets were applied to some indices (e.g., Mg1, Mg2),
while no corrections were needed for narrow-band indices such as
Mg$b$.
%
The correlations between different metal line indices are well defined
and give a consistent picture of metal line strengths in the M\,104
globular clusters. Analysis of the relative abundances of iron-peak
versus $\alpha$-elements, and comparison with models including
non-solar abundance ratios (e.g., \cite{thom+02}), will be presented in
a future paper.

The metallicities of globular clusters in M\,104 were derived
following the empirical approach based on the index-metallicity
relations for Galactic clusters.  Weighted mean metallicities were
calculated using several indices \cite{brod+huch90}. The assumptions
behind the empirical calibrations are that clusters in M\,104 are as
old as those in the Milky Way, and the abundance ratios follow the
same trend with metallicity as in the Milky Way globular cluster
system. This approach has the advantage of providing a robust,
model-independent ranking based on direct comparison of the line
strengths in the target clusters with those of the Milky Way
globulars.

\begin{figure}[t]
\begin{center}
\includegraphics[width=1.\textwidth]{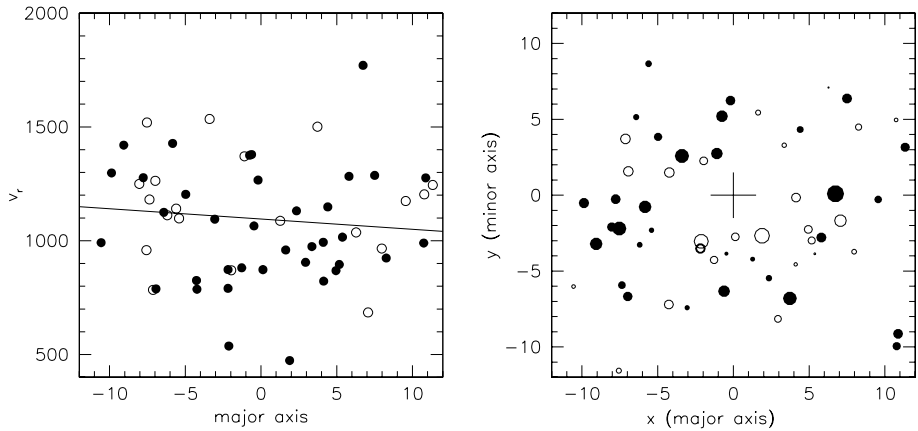}
\end{center}
\caption[]{{\it Left:} the radial velocity vs. radius for globular
clusters around the Sombrero galaxy. East is to the left.  The {\it
open circles} are globular clusters less metal-rich than
[Fe/H]$ = -1.2$. The {\it line} is a least squares fit to the data.
{\it Right:} the spatial distribution of clusters with approaching
({\it open circles}) and receding velocities ({\it filled
circles}). The size of the symbols is proportional to the velocity
relative to the systemic motion.  A {\it cross} marks the galaxy
center. }
\label{f_heldF3}
\end{figure}

For some indices, the index-metallicity relation
becomes distinctly non-linear for metal-rich clusters. An example is
given in Fig.~\ref{f_heldF1} for the Mg2 versus [Fe/H] relation.  
The non-linearity is confirmed by the spectral synthesis models, and was
previously noticed by others (\cite{greg94}, \cite{kiss+98}). 
As a result, a straightforward extrapolation of the original linear
calibrations above [Fe/H]$ \sim -0.7$ would seriously overestimate the
metallicity of the stronger-lined clusters.
As a simple approximation of the non-linear behavior of Mg2 and other
indices, we adopted two different slopes for clusters below and above
[Fe/H]$ = -0.7$. For clusters less metal-rich than this value, our
calibration is identical to the original one \cite{brod+huch90}.
Details on the new relations for very metal-rich clusters, as well as
abundances derived from comparisons with models, will be given 
elsewhere.
%
We find a range in metallicity from [Fe/H]$ = -2.13$ to [Fe/H]$ = +0.26$
with a mean value [Fe/H]$ = -0.90$. Some clusters with super-solar
abundances are indeed present in M\,104. Very few clusters appear to
be more metal-poor than [Fe/H]$ = -1.8$.

Diagnostic diagrams of Balmer line strengths against metal line
indices were used to estimate the ages of globular clusters in the
halo of M\,104.  Figure~\ref{f_heldF2} shows the distribution of
clusters in the diagram using H$\beta$ against Mg2.  Most globular
clusters in M\,104 appear consistent with an old age, comparable to
that of Milky Way globular clusters.  The clusters in the halo of
M\,104 appear to be coeval with the clusters in the inner (bulge)
regions studied by \cite{lars+02}.  Only for a few clusters is
H$\beta$ strong enough to be consistent with an age $\sim 2$
Gyr. Measurements of other Balmer line indices are in progress.

\section{Kinematics}

\begin{figure}[t]
\begin{center}
\includegraphics[width=1.\textwidth]{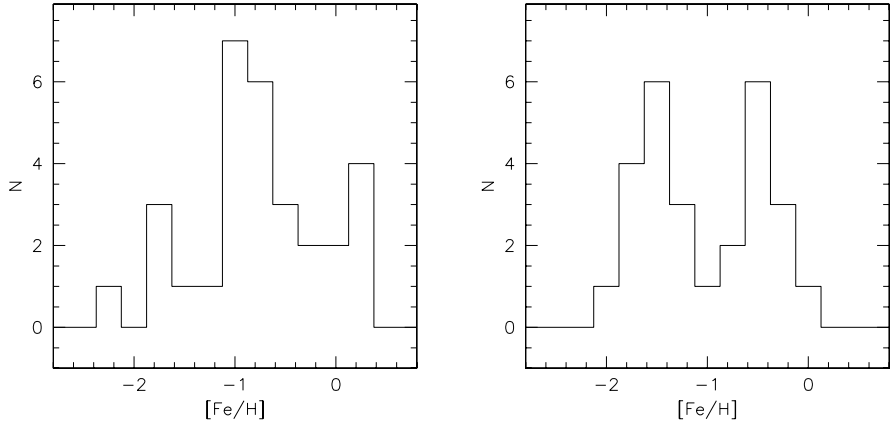}
\end{center}
\caption[]{The metallicity distribution of globular clusters in
M\,104, plotted separately for the inner ($r < 7$ kpc, {\it left
panel}) and outer regions ($r \ge 7$ kpc, {\it right}).  This figure
indicates a composite nature for the globular cluster system. }
\label{f_heldF4}
\end{figure}

Figure~\ref{f_heldF3} shows the radial velocities of candidates with
$400 < v_r < 2000$ \kms, which are probable members of the globular
cluster system of M\,104. The clusters on the E side have $\langle v_r
\rangle = 1132$ \kms\ with a velocity dispersion $\sigma_v = 187$
\kms.  On the W side, the velocity dispersion is $\sigma_v = 234$
\kms\ and the mean radial velocity 1052 \kms.  The velocity
distribution is non-gaussian on the E side due to the presence of a
distinct group of clusters with radial velocities $< 900$ \kms. 
Given the large scatter of the velocity measurements, the evidence for
rotation is marginal. A linear fit including all data points gives a
small difference for the mean cluster velocities on the E and W sides,
indicating a very shallow rotation curve, $\sim 4.5$ \kmskpc.
With the present data, no difference is obvious between the kinematics
of metal-rich and metal-poor clusters.  A larger sample of precise
radial velocities will be necessary to establish the presence of
rotation for the M\,104 halo clusters, and further observations are
planned to this aim.
Figure~\ref{f_heldF3} also shows (right panel) the spatial
distribution of clusters with approaching and receding velocity.  This
distribution might suggest that the kinematic axis is inclined with
respect to the photometric minor axis.

\section{Conclusions}

The metallicities derived from spectroscopy of globular clusters in
the bulge and halo of M\,104 confirm a bimodal [Fe/H] distribution, in
agreement with the results from photometry (\cite{lars+01}; Moretti et
al. 2002, this volume) and with the spectroscopic results for the
inner cluster sample \cite{lars+02}.  A two-gaussian fit to the
distribution has two peaks at [Fe/H]$ \approx -1.7$ and [Fe/H]$ \approx
-0.7$, very similar to those of globular clusters in our Galaxy (see
also \cite{lars+01}).
The abundance distribution indicates that out to 15 kpc, the
metal-rich clusters are 4 times more abundant than the metal-poor
ones.  A plot of globular cluster metallicities against the distance
from the center shows that metal-rich clusters are more numerous near
the galaxy center.
This radial gradient has been mostly interpreted as caused by a
changing mix of two (metal-rich and metal-poor) cluster populations.
However, Fig.~\ref{f_heldF4} suggests that the global [Fe/H]
distribution is produced by superposition of cluster populations whose
properties vary in a complex way as a function of distance from the
galaxy center.  Bimodality seems to be an (over)simplified description
of the metallicity distribution of globular clusters in M\,104 --
reality is probably more complex than we thought in this and many
other galaxies.



\begin{thebibliography}{8.}
\addcontentsline{toc}{section}{References}

\bibitem{bres+96}
A. Bressan, C. Chiosi, R. Tantalo: A\&A \textbf{311}, 425 (1996)

\bibitem{brid+97} 
T. J. Bridges, K. M. Ashman, S. E. Zepf, et al.: MNRAS \textbf{284}, 376 (1997)

\bibitem{brod+huch90}
J. P. Brodie, J. P. Huchra: ApJ \textbf{362}, 503 (1990)

\bibitem{greg94}
M. D. Gregg: AJ \textbf{108}, 2164 (1994)

\bibitem{kiss+98} M. Kissler-Patig, J. P. Brodie, L. L. Schroder, et
al.: AJ \textbf{115}, 105 (1998)

\bibitem{lars+01}
S. S. Larsen, D. A. Forbes, J. P. Brodie: MNRAS  \textbf{327}, 1116 (2001)

\bibitem{lars+02} S. S. Larsen, J. P. Brodie, M. A. Beasley,
D. A. Forbes: AJ \textbf{124}, 828 (2002)


\bibitem{thom+02} D. Thomas, C. Maraston, R. Bender: submitted to MNRAS 
(astro-ph/0209250) (2002)

\bibitem{vazd99}
A. Vazdekis: ApJ \textbf{513}, 224 (1999)

\bibitem{wort94}  G. Worthey, S. M. Faber, J. J. Gonzalez,
D. Burstein: ApJS \textbf{94}, 687 (1994)

\end{thebibliography}
\end{document}